\documentclass[12pt]{article}

\usepackage{amsfonts}
\usepackage{amsmath}
\usepackage{epsfig}
\usepackage{latexsym}
\usepackage{graphicx}
\numberwithin{equation}{section}
\allowdisplaybreaks

\usepackage[utf8]{inputenc}

\def\be{\begin{equation}}
\def\ee{\end{equation}}
\def\bea{\begin{eqnarray}}
\def\eea{\end{eqnarray}}
\newcommand{\G}{\Gamma}

\def\half{{1\over 2}}

\def\nn{\nonumber}

\renewcommand{\thefootnote}{\fnsymbol{footnote}}

\begin{document}


\hfuzz=100pt

\title{\begin{flushright}
\vspace{-3cm}
{\normalsize YITP-18-24}
\end{flushright}
\vspace{2cm}
{\Large \bf{Airy Function and 4d Quantum Gravity}}}
\date{}
\author{Pawe{\l} Caputa$^{a}$\footnote{pawel.caputa@yukawa.kyoto-u.ac.jp}\; and
Shinji Hirano$^{b, a}$\footnote{shinji.hirano@wits.ac.za}
  \\ \\ 
  $^a${\small{\it Center for Gravitational Physics,
Yukawa Institute for Theoretical Physics}}
\\{\small{\it Kyoto University, Kyoto 606-8502, Japan }}\\
$^b${\small{\it School of Physics and Mandelstam Institute for Theoretical Physics}}
\\{\small{\it \& DST-NRF Centre of Excellence in Mathematical and Statistical Sciences (CoE-MaSS) }}
\\{\small{\it University of the Witwatersrand, WITS 2050, Johannesburg, South Africa}}\\ 
\\
}
\date{}

\maketitle

\centerline{}

\begin{abstract}
We study four-dimensional quantum gravity with negative cosmological constant in the minisuperspace approximation and compute the partition function for the $S^3$ boundary geometry. 
In this approximation scheme the path integrals become dominated by a class of asymptotically AdS \lq\lq microstate geometries.''
Despite the fact that the theory is pure Einstein gravity without supersymmetry, the result precisely reproduces, up to higher curvature corrections, the Airy function in the $S^3$ partition function of the maximally supersymmetric Chern-Simons-matter (CSM) theory which sums up all perturbative $1/N$ corrections. We also show that this can be interpreted as a concrete realization of the idea that the CFT partition function is a solution to the Wheeler-DeWitt equation as advocated in the holographic renormalization group.
Furthermore, the agreement persists upon the inclusion of a string probe and it reproduces the Airy function in the vev of half-BPS Wilson loops in the CSM theory.
These results may suggest that the supergravity path integrals localize to the minisuperspace in certain cases and the use of the minisuperspace approximation in AdS/CFT may be a viable approach to study $1/N$ corrections to large $N$ CFTs.

\end{abstract}

\renewcommand{\thefootnote}{\arabic{footnote}}
\setcounter{footnote}{0}

\newpage

\section{Introduction}

One of the most well-known examples of Airy function in physics appears in the WKB approximation in which the Airy function bridges the wavefunctions across \lq\lq classical'' and \lq\lq quantum'' regimes of the potential. It also makes universal appearances in random matrix theory at the edge of eigenvalue distributions \cite{Tracy:1992rf, Forrester:1993vtx, 2011JSMTE..04..001N}. This is known as the Tracy-Widom distribution, and it sits at a crossover from the weak to strong coupling phase of some system and becomes a point of a 3rd order phase transition in the limit of large degrees of freedom. It thus seems that Airy functions tend to emerge at the boundary of two regimes. 
More recently, the Airy function made a surprising appearance in AdS/CFT correspondence \cite{Maldacena:1997re}.
In the case of the duality between ${\cal N}=6$ $U(N)_k\times U(N)_{-k}$ Chern-Simons-matter theory (ABJM theory) and type IIA string theory on $AdS_4\times \mathbb{C}P^3$ or M-theory on $AdS_4\times S^7/\mathbb{Z}_k$ \cite{Aharony:2008ug, Aharony:2008gk}, the $S^3$ partition function of ABJM theory turned out to be an Airy function   \cite{Fuji:2011km, Marino:2011eh}
\be
Z_{\rm ABJM}(S^3)\propto 
\mbox{Ai}\left[\left({\pi N^2\over\sqrt{2\lambda}}\right)^{2\over 3}\left(1-{1\over 24\lambda}-{\lambda\over 3N^2}\right)\right]
\label{allgenus}
\ee
where $\lambda=N/k$ is the 't Hooft coupling. This is the perturbative part of the full partition function in $1/N$ expansions and is supplemented by the two classes of nonperturbative tails  \cite{Marino:2011eh, Drukker:2011zy, Hatsuda:2012dt, Hatsuda:2013oxa}, somewhat similar to the Tracy-Widom distribution.\footnote{In the case of $k=1, 2$ when SUSY enhances to ${\cal N}=8$, the fully nonperturbative exact partition function has been found \cite{Codesido:2014oua}.} 
Importantly, via AdS/CFT, the Airy function \eqref{allgenus} corresponds to the all-loop perturbative quantum gravity partition function.
In some sense this occurrence of the Airy function is similar to WKB and Tracy-Widom in that it is the bridge between \lq\lq classical'' and nonperturbative tails. 

Our focus in this paper is to understand \eqref{allgenus} from the viewpoint of four-dimensional quantum gravity. 
Although it has been demonstrated that \eqref{allgenus} can be reproduced by the supergravity localization \cite{Dabholkar:2014wpa}, we wish, in particular, to show that the emergence of the Airy function is not all due to the sophistications of supergravity and extra dimensions but rather lies at the core of 4d quantum gravity. Motivated by the appearance of Airy functions in quantum cosmology \cite{Halliwell:1988wc, Halliwell:1988ik}, we study 4d quantum gravity with negative cosmological constant in the minisuperspace approximation to see if pure Einstein gravity suffices to reproduce the $S^3$ partition function of ABJM theory.
A mental image behind this idea is that stripping down gravity to the minisuperspace is likened to zooming into the edge of eigenvalue distributions in Tracy-Widom or to the turning points in WKB.
In this approximation scheme the path integrals become dominated by a class of asymptotically AdS \lq\lq microstate geometries'' and not only does the minisuperspace approximation reproduce the $S^3$ partition function but also the vev of half-BPS Wilson loops of ABJM theory corresponding to a string probe added to the Einstein gravity.
These results may suggest that the supergravity path integrals localize to the minisuperspace in certain cases and the use of the minisuperspace approximation in AdS/CFT may be a viable approach to study $1/N$ corrections to large $N$ CFTs.\footnote{It should be mentioned that our idea has a strong resemblance to that of \cite{Ooguri:2005vr}.}

The organization of our paper is as follows:
In Section \ref{MSS} we lay out the scheme of minisuperspace approximation adapted to the holographic setup.
In Section \ref{PT} we compute the partition function of 4d quantum gravity by explicitly performing path integrals in the minisuperspace approximation and show that it indeed reproduces the Airy function in the $S^3$ partition function of ABJM theory.
In Section \ref{WL}, in order to demonstrate that  the agreement extends beyond the $S^3$ partition function, we calculate the gravity partition function with a string probe and show that it indeed agrees with the vev of half-BPS Wilson loops in ABJM theory. 
In Section \ref{WDW} we clarify how the $S^3$ partition function is related to the \lq\lq wavefunction of the universe'' which solves the Wheeler-DeWitt equation and discuss the holographic RG interpretation of the wavefunction. 
In Section \ref{Discussions}, besides summarizing our results, we discuss further generalizations of our calculations. In particular, as an example, we present our result for the two point function of heavy operators. We also make brief comments on the positive cosmological constant case in light of our analysis performed for negative cosmological constant. 

\section{Path integrals in minisuperspace approximation}
\label{MSS}
In what follows, we are going to be interested in the partition function of 4d quantum gravity \cite{Gibbons:1976ue} dual to the CFT$_3$ sphere partition function at strong couplings in the duality between ABJM theory and type IIA string theory on $AdS_4\times \mathbb{C}P^3$. In principle, we should aim to perform the Euclidean path integral computation with the fully-fledged supergravity including higher curvature corrections. As we will show, however, the Airy function can be reproduced only from semiclassical path integrals of pure Einstein gravity with negative cosmological constant in the minisuperspace approximation. In this section we set up the Euclidean path integrals of pure Einstein gravity.

In the path integral approach to quantum gravity, we are to integrate over all hyperbolic Euclidean metrics with the $S^3$ boundary condition. In the ADM decomposition the general metrics can be parametrized as
\be
ds^2=N^2dr^2+\gamma_{\mu\nu}\left(dx^\mu+N^\mu dr\right)\left(dx^\nu+N^\nu dr\right),
\ee
where $r$ is the radial direction and $x^\mu$ are the coordinates of the 3d Euclidean space. 
Note that this is not the standard ADM decomposition in that the (Euclidean) time is replaced by the radial coordinate $r$. 
In other words, it is adapted to the holographic study in which Cauchy surfaces are timelike. 
Now, we decide for ourselves to work in the mimisuperspace approximation, and this will prove to be the crucial step for our analysis.  Namely, we restrict the space to be spherically symmetric:
\be
ds^2=N^2(r)dr^2+a^2(r)d\Omega^2_{3},\label{mSS}
\ee
where $N(r)$ is the laps function, $a(r)$ the scale factor and $d\Omega^2_{3}$ is the metric on $S^3$. For simplicity we will often omit the arguments of $N(r)$ and $a(r)$.

The path integrals require a very careful treatment that involves introducing ghosts (see e.g. \cite{Halliwell:1988wc}) even in the minisuperspace approximation. Fortunately, it is well-known that after these subtle steps one is left with path integrals over the laps $N(r)$ and the scale factor $a(r)$, 
\be
Z=\int DN \int Da\, e^{-S_E[N,a]} 
\ee
where 
\be
S_E[N,a]=S_{EH}+S_{GH}+S_{ct}
\ee
is the regularized finite action on Euclidean metrics which we elaborate on below.

The Euclidean action with negative cosmological constant is the standard 4d Einstein-Hilbert action with the Gibbons-Hawking-York boundary term \cite{Gibbons:1976ue, York:1972sj} (see Appendix \ref{Ddim} for details and conventions). For the metrics \eqref{mSS}, after we integrate over the $S^3$ coordinates, the action can be written as
\bea
S_{EH}+S_{GH}=-\frac{V_3}{8\pi G_N}\int dr N\left[3a\left(1+\frac{a'^2}{N^2}\right)-\Lambda a^3\right],
\eea
where $V_3$ is the area of the unit 3-sphere and the cosmological constant $\Lambda=-3/\ell^2$ with the AdS radius $\ell$.

The next important step is to transform the \lq\lq kinetic term'' of the scale factor into the canonical form. This is a well-known step in the Lorentzian de Sitter approach to the Hartle-Hawking wave function \cite{Hartle:1983ai, Halliwell:1988ik}, and it proves to be crucial in our discussion too. This transformation is done in two steps: First, we rescale the laps function $N\to N/a$ and then introduce a new variable $q=a^2$ so that the action becomes
\be
S_{EH}+S_{GH}=-\frac{3V_3}{8\pi G_N\ell^2}\int dr \left[\frac{\ell^2q'^2}{4N}+N\left(q+\ell^2\right)\right].\label{Actionq}
\ee
This is our main object in the following analysis and the gravitational path integrals are over the laps $N$ and $q$.\footnote{There is a subtlety in the choice of path integral measures. We will justify our choice of the measures a posteriori by requiring consistency with the Wheeler- DeWitt equation, as will be discussed in Section \ref{WDW}.}

Finally, in the computation of the partition function, we need to regularize the divergent terms by adding the counter-term action \cite{Balasubramanian:1999re, Emparan:1999pm, deHaro:2000vlm}. In four dimensions it is enough to add the local action with the scalar curvature of the induced metric and the cosmological constant term on the boundary  as in \eqref{CTa}. In the minisuperspace approximation it reads
\be
S_{ct}=\frac{V_3}{8\pi G_N \ell}\left(2q_{\infty}^{3/2}+ 3\ell^2q_{\infty}^{1/2}\right)\ ,\label{SCT}
\ee
where $q_{\infty}\gg \ell^2$ is the cutoff near the asymptotic boundary of the space.
With \eqref{Actionq} and \eqref{SCT}, we can now proceed to perform the path integrals over the laps $N$ and the redefined scale factor $q$.
\section{The $S^3$ partition function}
\label{PT}

The $S^3$ partition function of CFT in the large $N$ and strong coupling limits corresponds to the classical partition function of gravity on $H_4$, {\it i.e.} Euclidean $AdS_4$ with the $S^3$ boundary (times internal manifolds) \cite{Witten:1998qj, Gubser:1998bc}: 
\begin{align}
\lim_{\stackrel{N\to\infty}{\scriptscriptstyle \lambda\to\infty}}Z_{\rm CFT}(S^3)=\exp\left(-S_E\right)\biggr|_{\rm on-shell}
\label{classical}
\end{align}
where $\lambda$ is the 't Hooft coupling and the on-shell indicates that the action is evaluated on the Euclidean AdS$_4$ (times internal manifolds) with suitable regularization and renormalization \cite{Balasubramanian:1999re, Emparan:1999pm, deHaro:2000vlm}.
In the case of the duality between ABJM theory and type IIA string theory on $AdS_4\times \mathbb{C}P^3$ this has been explicitly checked by Drukker, Mari\~no and Putrov \cite{Drukker:2011zy, Marino:2011nm}.
In contrast to ${\cal N}=4$ SYM on $S^4$, the $S^3$ partition function of ABJM theory receives nontrivial $1/N$ corrections dual to quantum gravity all loop perturbative effects
and, remarkably, it sums up to an Airy function \cite{Fuji:2011km, Marino:2011eh}.
Furthermore, the Airy function has been reproduced as the quantum gravity partition function of M-theory on $AdS_4$ times Sasaki-Einstein seven manifolds by the supergravity localization computation \cite{Dabholkar:2014wpa}.

The goal of this section is to go beyond the large $N$/classical limit of \eqref{classical}. 
In particular, we wish to show that the appearance of the Airy function is not all due to the sophistications of supergravity and extra dimensions but rather lies at the core of 4d quantum gravity. 
As we will see, the path integrals of minimal 4d Einstein gravity in the minisuperspace approximation suffice to reproduce the Airy function  in the $S^3$ partition function of ABJM theory at strong couplings.

As we discussed in Section \ref{MSS}, the gravity partition function in the minisuperspace approximation takes the form
\begin{align}
Z_{\rm G}(S^3)=\int {\cal D}N\int {\cal D}q\exp\left[\frac{3V_3}{8\pi G_N\ell^2}\int dr \left(\frac{\ell^2q'^2}{4N}+N\left(q+\ell^2\right)\right)-S_{ct}\right]\ .
\end{align}
where we used $\Lambda=-3/\ell^2$ and  $S_{ct}$ is the counter terms at the boundary cutoff \eqref{SCT}. 
We now choose the gauge in which the lapse $N$ is constant. Our strategy is first to perform the $q$-integral in the saddle point approximation. Since the Lagrangian does not explicitly depend on \lq\lq time'' $r$, the saddle point equation is given by the \lq\lq energy'' conservation:
\begin{align}
E={\ell^2\over 4N^2} q'^2-q-\ell^2\ .
\end{align}
It is most convenient to parametrize $E=q_0-\ell^2$, and the saddle point equation becomes
\begin{align}
\pm {2N dr\over\ell} ={dq\over \sqrt{q+q_0}}\qquad\Longrightarrow\qquad q=-q_0+\left({N(r-r_0)\over\ell}\right)^2\ ,
\label{saddle}
\end{align}
where $r_0$ is an unphysical constant, corresponding to the origin of \lq\lq time'', which can be shifted away, whereas $q_0$ is a parameter which characterizes each saddle point. In other words, we have a series of saddle points labeled by $q_0$. To find the partition function we sum over all the saddle points by integrating over $q_0$. Put differently, we are summing over a class of asymptotically AdS \lq\lq microstate geometries'' specified by $q_0$, as we now elabotate.

A few remarks are in order: (1) 
The space at the saddle point of any $q_0$ is asymptotically Euclidean AdS$_4$ of radius $\ell$
\begin{align}
ds^2&={N^2 \over q(r)}dr^2 + q(r)d\Omega_3^2\quad\stackrel{q\gg \ell^2}{\longrightarrow}\quad  \ell^2d\eta^2+N^2e^{2\eta}d\Omega_3^2\ ,
\end{align}
where $\eta = \log\left[{r\over\ell}\right]\gg 1$ and $q\simeq (Nr/\ell)^2$. However, the space deviates from AdS$_4$ in the bulk in contrast to the classical limit. 
(2) The range of $q$ is taken to be $q\in [0, q_{\infty}]$ so that the entire space (within the boundary cutoff $q_{\infty}$) is covered. 
Although there is a conical singularity at $q=0$, it is harmless and admissible. 
As we will discuss in Section \ref{WDW}, when the space is terminated at some finite $q$, the partition function is the \lq\lq wavefunction of the universe'' $\Psi(q)$ which solves the Wheeler-DeWitt equation and can conceivably be interpreted as an IR cutoff of the dual CFT. 

Using the equation \eqref{saddle} with the upper sign, (minus) the saddle point action yields
\begin{align}
\hspace{-.3cm}
-S_{0}={3V_3 \over 8\pi G_N\ell}\int_{0}^{q_{\infty}} dq \left[\sqrt{q+q_0}- {q_0-\ell^2\over 2\sqrt{q+q_0}}\right]
={3V_3 \over 8\pi G_N\ell}\left[\frac{2}{3}q_{\infty}^{3\over 2}+\ell^2 q_{\infty}^{\half}+{1\over 3}q_0^{3\over 2}-\ell^2q_0^{\half}\right]
\ .\label{saddleaction}
\end{align} 
The divergent pieces are precisely cancelled by the counter terms $S_{ct}(q_{\infty})$ in \eqref{SCT}.
We thus find within our approximations that 
\begin{align}
Z_{\rm G}(S^3)\simeq\int dN\int {\cal D}Q\int [dq_0]\exp\left[{3V_3 \over 8\pi G_N\ell}\left({1\over 3}q_0^{3\over 2}-\ell^2q_0^{\half}\right)
+\frac{3V_3}{32\pi G_N}\int dr {Q'(r)^2\over N}\right]\ ,
\label{S3partitionFull}
\end{align}
where $Q(r)$ is the fluctuation of $q(r)$ about the saddle point. The integration measure $[dq_0]$ of the saddle point parameter $q_0$ will be determined shortly. It is more illuminating to introduce a new variable by a simple reparametrization
\be
q_0=\ell^2 a_0^2\ .
\ee
We now choose the measure $[dq_0]\propto da_0$ whose justification we will argue momentarily. Converting 4d Newton's constant $G_N$ and the AdS$_4$ radius $\ell$ into the parameters of ABJM theory, $N$ and $\lambda=N/k$, by
\be
{3V_3\ell^2 \over 8\pi G_N}={\pi N^2\over \sqrt{2\lambda}}\ ,
\label{parameters}
\ee
the Einstein gravity partition function yields
\begin{align}
Z_{\rm G}(S^3)\propto {1\over 2\pi i}\int_{\cal C} da_0\exp\left[{\pi N^2\over \sqrt{2\lambda}}\left({1\over 3}a_0^{3}-a_0\right)\right]
\propto {\rm Ai}\left[\left({\pi N^2\over \sqrt{2\lambda}}\right)^{2\over 3}\right]\ .
\label{S3partition}
\end{align}
This is precisely the Airy function that appears in the $S^3$ partition function of ABJM theory at large $\lambda$!\footnote{The $N$ in this equation is the rank $N$ of the gauge group of the dual CFT and should not be confused with the laps $N$ in \eqref{S3partitionFull}.}

A few remarks are in order: 
(1) As we will see in Section \ref{WDW}, this Airy function coincides with the \lq\lq wavefunction of the universe'' $\Psi(q)$ at $q=0$. This may lend further support on the choice of the measure $[dq_0]=da_0$ and the integration contour ${\cal C}$ in \eqref{S3partition}.
(2) The saddle point parameter $a_0$ is identified with the chemical potential $\mu$ of the grand partition function of ABJM theory \cite{Marino:2011eh}. 
(3) The integrations over $Q(r)$ and the laps $N$ are yet to be performed. Although they may yield additional factors of $N^2/\sqrt{\lambda}$, they are secondary to our main point and may well be cancelled in supersymmetric cases.\footnote{We stress again that the $N$ of $N^2/\sqrt{\lambda}$ is the rank $N$ of the gauge group of the dual CFT and should not be confused with the laps $N$ in \eqref{S3partitionFull}.}
(4) The shift by $-1/(24\lambda)-\lambda/(3N^2)$ in the Airy function of ABJM theory \eqref{allgenus} cannot be accounted for in our approximations, since they originate from higher curvature corrections \cite{Bergman:2009zh, Aharony:2009fc}.

This result may suggest that the supergravity path integrals localize to the minisuperspace in certain cases. However, it is not clear how exactly this can be related to the work of Dabholkar, Drukker and Gomes \cite{Dabholkar:2014wpa}.


\section{A string probe and Wilson loops}
\label{WL}

One might wonder if the above agreement of the $S^3$ partition function is a mere coincidence. 
In order to argue that this may be more than a luck and to see how useful our approach might be, we shall show that the agreement extends beyond the partition function to half-BPS Wilson loops which have been calculated in ABJM theory \cite{Klemm:2012ii}. 

The BPS Wilson loops are also given in terms of Airy functions. In particular, the half-BPS Wilson loops of winding number $n$ take a simple form 
\be
\langle W_n^{1/2}\rangle\propto {\rm Ai}\left[\left({\pi N^2\over\sqrt{2\lambda}}\right)^{{2\over 3}}\left(1-{2n\lambda\over N^2}-{1\over 24\lambda}-{\lambda\over 3N^2}\right)\right]\ ,
\ee
where the piece $-{1\over 24\lambda}-{\lambda\over 3N^2}$ is the aforementioned shift originating from higher curvature corrections and cannot be captured by our approach. However, the contribution proportional to the winding number $n$ is dual
to an $n$-wound string \cite{Maldacena:1998im, Rey:1998ik, Drukker:2008zx} 
and corresponds to a simple shift in the coefficient of $a_0$ in \eqref{S3partition}. 
As we will show, a string probe of winding number $n$ precisely yields the right amount of shift to the coefficient.

In the gravity partition function with a string probe, the Nambu-Goto action is added to the saddle point action \eqref{saddleaction}:
\be
S_0(q_0)\to S_0(q_0) +S_{NG}(q_0)
\ee
where the NG action is the minimal surface bounded by a circular Wilson loop wrapping the great circle of the boundary $S^3$ embedded in the space 
\be
ds^2={N^2\over q(r)}dr^2 + q(r) d\Omega_3^2
\ee
 at the saddle point \eqref{saddle}.
Choosing the worldsheet coordinates to be $(r, \phi)$ with $\phi$ parametrizing the great circle of $S^3$, the induced metric on the string yields
\be
ds_{ws}^2={N^2\over q(r)}dr^2+q(r)d\phi^2\qquad\Longrightarrow\qquad
\det g_{ws}= N^2\ .
\ee
Here $N$ is the laps not to be confused with the rank $N$ of the CFT. 
This is trivially a minimal surface. In other words, our choice of the worldsheet coordinates happens to be the one naturally parametrizing the minimal surface. Note that this minimal surface corresponds to the AdS$_2$ in the classical case  \cite{Drukker:2008zx}.

The NG action for the $n$-wound string is then given by
\begin{align}
S_{NG}(q_0)=nT\int \!dr\!\int_0^{2\pi} d\phi\sqrt{\det g_{ws}}=\pi nT\ell\int_0^{q_{\infty}}\!{dq\over \sqrt{q+q_0}}=2\pi nT\ell\left[\sqrt{q_{\infty}+q_0}-q_0^{\half}\right]
\end{align}
where $T$ is the string tension and we used the saddle point equation \eqref{saddle}. 
Subtracting the divergent part, the total saddle point action is found to be
\begin{align}
S_{tot}(q_0)=-{3V_3 \over 8\pi G_N\ell}\left[{1\over 3}q_0^{3\over 2}-\left(1-{16\pi^2G_N n T\over 3V_3}\right)\ell^2q_0^{\half}\right]\ .
\end{align}
Using the relation among parameters $T\ell^2=\sqrt{\lambda\over 2}$ \cite{Drukker:2008zx}, this becomes
\be
S_{tot}(a_0)=-{\pi N^2\over \sqrt{2\lambda}}\left[{1\over 3}a_0^{3}-\left(1-{2n\lambda\over N^2}\right)a_0\right]\ .
\ee
As advertized, we thus find that the gravity partition function with an $n$-wound string probe precisely agrees with the half-BPS Wilson loops in ABJM theory:
\begin{align}
Z_{G+{\rm string}}(S^3)\propto  {1\over 2\pi i}\int_{\cal C} da_0e^{{\pi N^2\over \sqrt{2\lambda}}\left[{1\over 3}a_0^{3}-\left(1-{2n\lambda\over N^2}\right)a_0\right]}
\propto {\rm Ai}\left[\left({\pi N^2\over \sqrt{2\lambda}}\right)^{2\over 3}\left(1-{2n\lambda\over N^2}\right)\right]\ .
\label{halfBPSWL}
\end{align}
This agreement may bolster our claim that the supergravity path integrals may localize to the minisuperspace in certain cases and the use of the minisuperspace approximation in AdS/CFT may be a viable approach to study $1/N$ corrections to large $N$ CFTs.


\section{The Wheeler-DeWitt equation and RG flow}
\label{WDW}

It is expected that the partition function is a solution to the Wheeler-DeWitt equation. This is actually an alternative and much simpler way to find the partition function. However, it is not as obvious as it may seem how exactly the \lq\lq wavefunction of the universe'' can be identified with the $S^3$ partition function of the CFT.

The Hamiltonian constraint of \eqref{Actionq} yields the Wheeler-DeWittt equation (see \eqref{WDWdDim})
\begin{align}
\left[{d^2\over dq^2}-{9\pi^2\over 16G_N^2\ell^2}\left(q+\ell^2\right) \right]\Psi(q)=0\ ,
\end{align}
where we used the canonical momentum $\pi_q=\hbar{d\over dq}$ with $\hbar=1$. Note that since the \lq\lq time'' is the spatial radial coordinate $r$ and can be regarded as a Euclidean time, the imaginary $i$ is absent in  $\pi_q=\hbar{d\over dq}$.
This is the Airy equation and can be solved to 
\begin{align}
\Psi(q)=C_1{\rm Ai}\left[\left({3\pi\ell^2\over 4G_N}\right)^{2\over 3}\left(\ell^{-2}q+1\right)\right]+C_2{\rm Bi}\left[\left({3\pi\ell^2\over 4G_N}\right)^{2\over 3}\left(\ell^{-2}q+1\right)\right]\ .
\end{align}
By using \eqref{parameters} we have ${3\pi\ell^2\over 4G_N}={\pi N^2\over\sqrt{2\lambda}}$ and observe that 
\be
Z_G(S^3)\propto \Psi(0)
\ee
with the choice $C_2=0$. Thus this provides a concrete realization of the idea that the CFT partition function is a solution to the WDW equation as advocated in the holographic renormalization group \cite{deBoer:1999tgo, McGough:2016lol}.

In order to understand the relation between $\Psi(q)$ and the partition function for generic $q$, we go back to the saddle point action \eqref{saddleaction} and terminate the space at a finite $q$ instead of going all the way down to $q=0$:
\begin{align}
-S_0\quad\to\quad -S_{0}(q)&={3V_3 \over 8\pi G_N\ell}\int_{q}^{q_{\infty}} dq' \left[\sqrt{q'+q_0}- {q_0-\ell^2\over 2\sqrt{q'+q_0}}\right]\nn\\
&={3V_3 \over 8\pi G_N\ell}\left[\frac{2}{3}q_{\infty}^{3\over 2}+\ell^2 q_{\infty}^{\half}+{1\over 3}Q_0^{3\over 2}-(q+\ell^2)Q_0^{\half}\right]
\ ,
\end{align} 
where we introduced the shifted parameter $Q_0=q_0+q$. Integrating over $Q_0$ in the partition function, we find that
\begin{align}
Z_{\rm G}(S^3;q)\propto {\rm Ai}\left[\left({\pi N^2\over \sqrt{2\lambda}}\right)^{2\over 3}\left(\ell^{-2}q+1\right)\right]
\propto \Psi(q)\ .
\label{cutoffS3partition}
\end{align}
Since the radial scale $q$ corresponds to the energy scale of the CFT, it is most natural to interpret $q$ as the IR cutoff in the CFT and $\Psi(q)$ as the IR cutoff $S^3$ partition function in which only the modes above the energy scale $q$ are integrated out.
It is curious to observe that at large $N$ the \lq\lq free energy'' $F=-\ln|\Psi(q)|$ monotonically decreases as the IR scale $q$ decreases in accordance with the F-theorem proposed in \cite{Jafferis:2011zi}.\footnote{The RG flow here is not driven by some relevant operators but rather ad hoc and induced by a hard wall at the IR cutoff $q$ as in \cite{Polchinski:2001tt, Erlich:2005qh}. We thank Tadashi Takayagi for his comment concerning this point.}

We note that this discussion may also lend support on the justification of the choice of path integral measures assumed in preceding sections.

\section{Discussions and conclusions}\label{Discussions}

We studied 4d quantum gravity with negative cosmological constant in the minisuperspace approximation and computed the partition function with or without a string probe. 
In this approximation scheme the path integrals become dominated by a class of asymptotically AdS \lq\lq microstate geometries.''
Despite the fact that the theory is pure Einstein gravity without supersymmetry, the results precisely reproduce, up to higher curvature corrections, the Airy functions in the $S^3$ partition function and vev of half-BPS Wilson loops of ABJM theory, which sums up all $1/N$ corrections and corresponds, via AdS/CFT, to the all-loop perturbative quantum gravity result. 

We  would like to see how viable this approach actually is for studying $1/N$ corrections and how far it can be pushed. As an immediate application,  
for example, it is straightforward to generalize our computation to two point functions of heavy operators for which the geodesic approximation of a heavy particle probe suffices. We only quote the final result:
\begin{align}
\langle{\cal O}_J(\Delta\theta/2){\cal O}_J(-\Delta\theta/2)\rangle_{S^3}\propto {1\over 2\pi i}\int_{\cal C} da_0
\left({\ell a_0 \over \sin{\Delta\theta a_0\over 2}}\right)^{2J}
\exp\left[{\pi N^2\over\sqrt{2\lambda}}\left({1\over 3}a_0^3-a_0\right)\right]\ ,
\label{2pt}
\end{align}
where $\Delta\theta$ is the latitude distance between the two operators on $S^3$ and $J$ is the dimension of the operators ${\cal O}_J$ and related to the mass $m$ of the particle by $\ell m =J \gg 1$. At large $N$ the leading correction to the 2pt function normalization can be found as $\exp(-\sqrt{2\lambda}J^2/(\pi N^2))$ which corresponds to the correction from one-loop Witten diagrams. To our knowledge, however, we currently lack the data on the 3d CFT side to make comparisons. Instead, we might regard the minisuperspace two point function \eqref{2pt} as a prediction.

In the case of four-dimensional gravity the Ryu-Takayanagi (RT) minimal surface for the entanglement entropy \cite{Ryu:2006bv} is two-dimensional and technically coincides with the minimal surface of Wilson loops apart from the overall constant. When our Wilson loop result is translated to the holographic entanglement entropy, its finite part reads
\begin{align}
S_{\rm HEE}({\rm equator})= {1\over 4G_N}\ln\left[
 {\rm Ai}\left[\left({3\pi \ell^2\over 4G_N}\right)^{2\over 3}\right]/{\rm Ai}\left[\left({3\pi \ell^2\over 4G_N}\right)^{2\over 3}\left(1-{8G_N \over 3\ell^2}\right)\right]\right]\ .
\end{align}
This is purportedly a RT minimal surface area (divided by $4G_N$) corrected by quantum gravity effects and our approach might provide a tool to test the proposed interpretation of quantum corrections given in \cite{Faulkner:2013ana}. 
It would be interesting to gereralize this result to the entanglement between the hemi-spheres divided at a generic latitude line.

One of the most interesting applications is to the case with the $S^1\times S^2$ boundary geometry and study $1/N$ corrections to the black hole entropy. In the large $N$ limit the precise agreement was found between the gravity and ABJM theory computations \cite{Benini:2015eyy}. It would be very interesting to see whether our minisuperspace approach can correctly reproduce $1/N$ corrections of the ABJM index result.
Another interesting application is to study geodesic Witten diagrams \cite{Hijano:2015zsa} and find general structures of $1/N$ corrections to conformal blocks.  

We also discussed how exactly the $S^3$ partition function is related to the \lq\lq wavefunction of the universe'' $\Psi(q)$ which solves the Wheeler-DeWitt equation and showed that it is in fact the wavefunction $\Psi(q)$ at $q=0$. This can then be interpreted as a concrete realization of the idea that the CFT partition function is a solution to the Wheeler-DeWitt equation as advocated in the holographic renormalization group \cite{deBoer:1999tgo, McGough:2016lol}.
Given the relation of $q$ to the IR cutoff of the CFT, we proposed how the wavefunction $\Psi(q)$ for generic $q$ can be interpreted in the CFT.

We also note that since the Airy function was found by solving the holomorphic anomaly equation (HAE) \cite{Bershadsky:1993ta} in the original derivation of \cite{Fuji:2011km}, it is natural to ask whether the WDW equation can possibly be identified with the HAE in any way.  
However, since the two equations are rather different in concepts and technical details, it is not obvious if and how they can be identified. 
Nevertheless, it is worth pointing out that a speculation was made in  \cite{Witten:1993ed} that the HAE might have an interpretation as the WDW equation.

Finally, we would like to make comments on the de Sitter case and Maldacena's proposal on the relation between the $S^3$ partition function and the Hartle-Hawking measure \cite{Maldacena:2002vr, Maldacena:2011mk}. A similar calculation yields the partition/wave function for the dS case
\begin{align}
Z_{dS}(S^3; q)\propto {1\over 2\pi i}\int_{{\cal C}_{dS}} da_0\exp\left[{3\pi\ell_{dS}^2 \over 4 G_N}\left({1\over 3}a_0^{3}-(1-\ell_{dS}^{-2}q)a_0\right)\right]
\end{align}
which is an Airy function and a solution to the WDW equation. For the HH measure \cite{Hartle:1983ai} the integration contour ${\cal C}_{dS}$ must be chosen such that the partition function has the exponentially growing component, ${\rm Bi}[(3\pi\ell_{dS}^2/(4G_N))^{2/3}(1-\ell_{dS}^{-2}q)]$. In the classical limit or the saddle point approximation to the $a_0$ integration with the HH boundary, the dS and AdS results are related by the simple analytic continuation $\ell_{dS}^2\to -\ell_{AdS}^2$ \cite{Maldacena:2002vr, Maldacena:2011mk}.
However, in the quantum case, since the continuation crosses a Stokes and an anti-Stokes line,  care is needed to pick up the subdominant contribution. 

\section*{Acknowledgment}

We would like to thank Sumit R. Das, Sanefumi Moriyama, Masaki Shigemori, Tadashi Takayanagi and Ali Zahabi for discussions and comments. 
SH would like to thank the Yukawa Institute for Theoretical Physics and the Graduate School of Mathematics at Nagoya University for their kind hospitality.  The work of PC was supported by the Simons Foundation through the ``It from Qubit'' collaboration and by the JSPS starting grant KAKENHI 17H06787.  
The work of SH was supported in part by the National Research Foundation of South Africa and DST-NRF Centre of Excellence in Mathematical and Statistical Sciences (CoE-MaSS).
Opinions expressed and conclusions arrived at are those of the author and are not necessarily to be attributed to the NRF or the CoE-MaSS.

\appendix
\renewcommand{\theequation}{\Alph{section}.\arabic{equation}}

\section{Details and conventions}\label{Ddim}
For completeness, we present a detailed derivation of the gravity action in the minisuperspace approximation as well as the WDW equation used in the main text for arbitrary dimensions $d+1$.\\
The Euclidean gravity action is defined as
\be
S_{EH}+S_{GH}=-\frac{1}{16\pi G_N}\int_{\mathcal{M}}d^{d+1}x\sqrt{g}\left(R-2\Lambda\right)+\frac{1}{8\pi G_N}\int_{\partial \mathcal{M}}d^dx\sqrt{\gamma}\Theta
\ee
with the negative the cosmological constant 
\be
\Lambda=-\frac{d(d-1)}{2\ell^2}\ .
\ee
The minisuperspace ansatz for the metric is given by
\be
ds^2=g_{\mu\nu}dx^\mu dx^\nu=N^2(r)dr^2+a^2(r)d\Omega^2_{d}\ ,
\ee
where $d\Omega^2_{d}$ is a metric on the $d$-dimensional sphere with the volume
\be
V_{d}=\int d\Omega_d=\frac{2\pi^{\frac{d+1}{2}}}{\G\left(\frac{d+1}{2}\right)}\ .
\ee
The Ricci scalar can be expressed in terms of the laps and the scale factor as 
\be
R=d(d-1)\left[\frac{1}{a^2(r)}-\frac{a'(r)^2}{a^2(r)N^2(r)}\right]+2d\left[\frac{a'(r)N'(r)}{a(r)N(r)^3}-\frac{a''(r)}{a(r)N^2(r)}\right].
\ee
Meanwhile, the extrinsic curvature is defined by
\be
\Theta^{\mu\nu}=-\frac{1}{2}\left(\nabla^\mu\hat{n}^\nu+\nabla^\nu\hat{n}^\mu\right)\ ,
\ee
and for the boundary at constant $r$ we have the normal vectors
\be
\hat{n}^\mu=N^{-1}(r)\delta^{\mu,r}\ ,\qquad g_{\mu\nu}\hat{n}^\mu\hat{n}^\nu=1
\ee
so that
\be
\Theta=-g_{\mu\nu}\nabla^\mu\hat{n}^\nu=\frac{N'(r)}{N^2(r)}-\G^\mu_{\mu r}N^{-1}(r)=-\frac{d\,a'(r)}{N(r)a(r)}\ ,
\ee
where we used the nonvanishing components of the Christoffel symbols
\be
\G^r_{rr}=\frac{N'(r)}{N(r)}\ ,\qquad \G^{\theta_i}_{\theta_i r}=\frac{a'(r)}{a(r)}\ .
\ee
The Einstein-Hilbert action then becomes
\bea
&&-\frac{1}{16\pi G_N}\int_{\mathcal{M}}d^{d+1}x\sqrt{g}(R-2\Lambda)=-\frac{V_d}{16\pi G_N}\int dr N(r)\left[d(d-1)a^{d-2}(r)\left(1+\frac{a'(r)^2}{N^2(r)}\right)\right]\nn\\
&&+\frac{V_d}{8\pi G_N}\int dr N(r) \Lambda a^d(r)+\frac{V_d d}{8\pi G_N}\int dr \partial_r\left(a^{d-1}(r)\frac{a'(r)}{N(r)}\right)\ .
\eea
On the other hand, the Gibbons-Hawking-York boundary term \cite{Gibbons:1976ue, York:1972sj} 
\be
\frac{1}{8\pi G_N}\int_{\partial \mathcal{M}}d^dx\sqrt{\gamma}\,\Theta=-\frac{V_d d}{8\pi G_N}\left[ \frac{a^{d-1}(r) a'(r)}{N(r)}\right]_{bdr}
\ee
precisely cancels the boundary contribution from the bulk action and we have
\be
S_{EH}+S_{GH}=-\frac{V_d}{8\pi G_N}\int dr N\left[\frac{d(d-1)}{2}a^{d-2}\left(1+\frac{a'^2}{N^2}\right)-\Lambda a^d\right].
\ee
Next, for the canonical kinetic term, we first redefine the laps function $N\to Na^{d-4}$ and the introduce a new variable $q=a^2$ that brings us to\\
\be
S_{EH}+S_{GH}=-\frac{V_d}{8\pi G_N}\int dr \left[\frac{d(d-1)}{2}\frac{q'^2}{4N}+N\left(\frac{d(d-1)}{2}q^{d-3}-\Lambda q^{d-2}\right)\right].\label{ActionqDd}
\ee\\
For $d=3$ this reproduces the action used in the main text.\\

To subtract the divergences we use the standard counter-term action \cite{Balasubramanian:1999re, Emparan:1999pm, deHaro:2000vlm}\footnote{This counter-term action is valid for $d=2,3,4$ i.e. $AdS_{3,4,5}$ and for $d=2$ i.e. $AdS_3$ we only take the first term.}
\be
S_{ct}=\frac{1}{8\pi G_N}\int_{\partial M}\sqrt{\gamma}\left(\frac{d-1}{\ell}+\frac{\ell}{2(d-2)} R_{c}(r)\right)\label{CTa}
\ee
where $\sqrt{\gamma}=a^d(r)d\Omega_d$ and the Ricci scalar of the induced metric at constant $r$ is
\be
R_c(r)=\frac{d(d-1)}{a^2(r)}.
\ee

Finally, we derive the Wheeler-DeWitt equation from \eqref{ActionqDd}: We first define the canonical \lq\lq momentum'' conjugate to $q(r)$
\be
p\equiv \frac{\partial L}{\partial q'}=-\frac{V_d}{8\pi G_N}\frac{d(d-1)}{4N}q'\ .
\ee
By the Legendre transformation $H=q'p-L$, we find the \lq\lq Hamiltonian'' 
\be
H=N\hat{H}=-\frac{16\pi G_N }{V_d d(d-1)}N\left[p^2-\left(\frac{d(d-1)V_d}{16\pi G_N\ell}\right)^2\left(\ell^2q^{d-3}+q^{d-2}\right)\right].
\ee
By using the differential form of the momentum, $p=\hbar\frac{d}{dq}$, we arrive at the Hamiltonian constraint, or the Wheeler-DeWitt equation, for the wavefunction
\be
\hat{H}\Psi(q)=\left[\hbar^2\frac{d^2}{dq^2}-\left(\frac{d(d-1)V_d}{16\pi G_N\ell}\right)^2\left(\ell^2q^{d-3}+q^{d-2}\right)\right]\Psi(q)=0\ .\label{WDWdDim}
\ee
In four dimensions ($d=3$), this becomes the Airy equation. It is also intriguing to note that in 5 dimensions ($d=4$) the equation can be written in the form of the Schr\"odinger equation for a simple harmonic oscillator whose solution is given in terms of Hermite polynomials.


\begin{thebibliography}{40}

\bibitem{Tracy:1992rf} 
  C.~A.~Tracy and H.~Widom,
  ``Level spacing distributions and the Airy kernel,''
  Commun.\ Math.\ Phys.\  {\bf 159}, 151 (1994)
  doi:10.1007/BF02100489
  [hep-th/9211141].
  
\bibitem{Forrester:1993vtx} 
  P.~J.~Forrester,
  ``The spectrum edge of random matrix ensembles,''
  Nucl.\ Phys.\ B {\bf 402}, 709 (1993).
  doi:10.1016/0550-3213(93)90126-A.
  
 \bibitem{2011JSMTE..04..001N} C.~Nadal and S.~N.~Majumdar, 
 ``A simple derivation of the Tracy-Widom distribution of the maximal eigenvalue of a Gaussian unitary random matrix,''
 J. Stat. Mech. (2011) P04001. arXiv:1102.0738 [cond-mat.stat-mech];
 S.~N.~Majumdar and G.~Schehr,
``Top eigenvalue of a random matrix: large deviations and third order phase transition,''
J. Stat. Mech. (2014) P01012. arXiv:1311.0580 [cond-mat.stat-mech]. 

\bibitem{Maldacena:1997re}
  J.~M.~Maldacena,
  ``The large $N$ limit of superconformal field theories and supergravity,''
  Adv.\ Theor.\ Math.\ Phys.\  {\bf 2}, 231 (1998)
  [Int.\ J.\ Theor.\ Phys.\  {\bf 38}, 1113 (1999)]
  [arXiv:hep-th/9711200].

\bibitem{Aharony:2008ug} 
  O.~Aharony, O.~Bergman, D.~L.~Jafferis and J.~Maldacena,
  ``${\cal N}=6$ superconformal Chern-Simons-matter theories, M2-branes and their gravity duals,''
  JHEP {\bf 0810}, 091 (2008)
  doi:10.1088/1126-6708/2008/10/091
  [arXiv:0806.1218 [hep-th]].
  
\bibitem{Aharony:2008gk} 
  O.~Aharony, O.~Bergman and D.~L.~Jafferis,
  ``Fractional M2-branes,''
  JHEP {\bf 0811}, 043 (2008)
  doi:10.1088/1126-6708/2008/11/043
  [arXiv:0807.4924 [hep-th]].


\bibitem{Fuji:2011km} 
  H.~Fuji, S.~Hirano and S.~Moriyama,
  ``Summing Up All Genus Free Energy of ABJM Matrix Model,''
  JHEP {\bf 1108}, 001 (2011)
  doi:10.1007/JHEP08(2011)001
  [arXiv:1106.4631 [hep-th]].
  
\bibitem{Marino:2011eh} 
  M.~Mari\~no and P.~Putrov,
  ``ABJM theory as a Fermi gas,''
  J.\ Stat.\ Mech.\  {\bf 1203}, P03001 (2012)
  doi:10.1088/1742-5468/2012/03/P03001
  [arXiv:1110.4066 [hep-th]].

\bibitem{Drukker:2011zy} 
  N.~Drukker, M.~Mari\~no and P.~Putrov,
  ``Nonperturbative aspects of ABJM theory,''
  JHEP {\bf 1111}, 141 (2011)
  doi:10.1007/JHEP11(2011)141
  [arXiv:1103.4844 [hep-th]].

\bibitem{Hatsuda:2012dt} 
  Y.~Hatsuda, S.~Moriyama and K.~Okuyama,
  ``Instanton Effects in ABJM Theory from Fermi Gas Approach,''
  JHEP {\bf 1301}, 158 (2013)
  doi:10.1007/JHEP01(2013)158
  [arXiv:1211.1251 [hep-th]];
  ``Instanton Bound States in ABJM Theory,''
  JHEP {\bf 1305}, 054 (2013)
  doi:10.1007/JHEP05(2013)054
  [arXiv:1301.5184 [hep-th]].
  
\bibitem{Hatsuda:2013oxa} 
  Y.~Hatsuda, M.~Mari\~no, S.~Moriyama and K.~Okuyama,
  ``Non-perturbative effects and the refined topological string,''
  JHEP {\bf 1409}, 168 (2014)
  doi:10.1007/JHEP09(2014)168
  [arXiv:1306.1734 [hep-th]].

\bibitem{Codesido:2014oua} 
  S.~Codesido, A.~Grassi and M.~Mari\~no,
  ``Exact results in ${\cal N}=8$ Chern-Simons-matter theories and quantum geometry,''
  JHEP {\bf 1507}, 011 (2015)
  doi:10.1007/JHEP07(2015)011
  [arXiv:1409.1799 [hep-th]].
  
\bibitem{Dabholkar:2014wpa} 
  A.~Dabholkar, N.~Drukker and J.~Gomes,
  ``Localization in supergravity and quantum AdS$_4$/CFT$_3$ holography,''
  JHEP {\bf 1410}, 90 (2014)
  doi:10.1007/JHEP10(2014)090
  [arXiv:1406.0505 [hep-th]].
 

\bibitem{Halliwell:1988wc} 
  J.~J.~Halliwell,
  ``Derivation of the Wheeler-De Witt Equation from a Path Integral for Minisuperspace Models,''
  Phys.\ Rev.\ D {\bf 38}, 2468 (1988).
  doi:10.1103/PhysRevD.38.2468.
  
\bibitem{Halliwell:1988ik} 
  J.~J.~Halliwell and J.~Louko,
  ``Steepest Descent Contours in the Path Integral Approach to Quantum Cosmology. 1. The De Sitter Minisuperspace Model,''
  Phys.\ Rev.\ D {\bf 39}, 2206 (1989).
  doi:10.1103/PhysRevD.39.2206.


\bibitem{Ooguri:2005vr} 
  H.~Ooguri, C.~Vafa and E.~P.~Verlinde,
  ``Hartle-Hawking wave-function for flux compactifications,''
  Lett.\ Math.\ Phys.\  {\bf 74}, 311 (2005)
  doi:10.1007/s11005-005-0022-x
  [hep-th/0502211].

\bibitem{Gibbons:1976ue} 
  G.~W.~Gibbons and S.~W.~Hawking,
  ``Action Integrals and Partition Functions in Quantum Gravity,''
  Phys.\ Rev.\ D {\bf 15}, 2752 (1977).
  doi:10.1103/PhysRevD.15.2752.
  
\bibitem{York:1972sj} 
  J.~W.~York, Jr.,
  ``Role of conformal three geometry in the dynamics of gravitation,''
  Phys.\ Rev.\ Lett.\  {\bf 28}, 1082 (1972).
  doi:10.1103/PhysRevLett.28.1082.
  
\bibitem{Hartle:1983ai} 
  J.~B.~Hartle and S.~W.~Hawking,
  ``Wave Function of the Universe,''
  Phys.\ Rev.\ D {\bf 28}, 2960 (1983).
  doi:10.1103/PhysRevD.28.2960.

\bibitem{Balasubramanian:1999re}
  V.~Balasubramanian and P.~Kraus,
  ``A Stress tensor for Anti-de Sitter gravity,''
  Commun.\ Math.\ Phys.\  {\bf 208} (1999) 413
  [hep-th/9902121].

\bibitem{Emparan:1999pm} 
  R.~Emparan, C.~V.~Johnson and R.~C.~Myers,
  ``Surface terms as counterterms in the AdS/CFT correspondence,''
  Phys.\ Rev.\ D {\bf 60}, 104001 (1999)
  doi:10.1103/PhysRevD.60.104001
  [hep-th/9903238].
  
\bibitem{deHaro:2000vlm} 
  S.~de Haro, S.~N.~Solodukhin and K.~Skenderis,
  ``Holographic reconstruction of space-time and renormalization in the AdS/CFT correspondence,''
  Commun.\ Math.\ Phys.\  {\bf 217}, 595 (2001)
  doi:10.1007/s002200100381
  [hep-th/0002230].
  
\bibitem{Witten:1998qj} 
  E.~Witten,
  ``Anti-de Sitter space and holography,''
  Adv.\ Theor.\ Math.\ Phys.\  {\bf 2}, 253 (1998)
  doi:10.4310/ATMP.1998.v2.n2.a2
  [hep-th/9802150].
  
\bibitem{Gubser:1998bc} 
  S.~S.~Gubser, I.~R.~Klebanov and A.~M.~Polyakov,
  ``Gauge theory correlators from noncritical string theory,''
  Phys.\ Lett.\ B {\bf 428}, 105 (1998)
  doi:10.1016/S0370-2693(98)00377-3
  [hep-th/9802109].

\bibitem{Drukker:2010nc} 
  N.~Drukker, M.~Mari\~no and P.~Putrov,
  ``From weak to strong coupling in ABJM theory,''
  Commun.\ Math.\ Phys.\  {\bf 306}, 511 (2011)
  doi:10.1007/s00220-011-1253-6
  [arXiv:1007.3837 [hep-th]].

\bibitem{Marino:2011nm} 
  M.~Mari\~no,
  ``Lectures on localization and matrix models in supersymmetric Chern-Simons-matter theories,''
  J.\ Phys.\ A {\bf 44}, 463001 (2011)
  doi:10.1088/1751-8113/44/46/463001
  [arXiv:1104.0783 [hep-th]].
    
\bibitem{Bergman:2009zh}
  O.~Bergman, S.~Hirano,
  ``Anomalous radius shift in AdS$_4$/CFT$_3$,''
  JHEP {\bf 0907}, 016 (2009).
  [arXiv:0902.1743 [hep-th]].

\bibitem{Aharony:2009fc}
  O.~Aharony, A.~Hashimoto, S.~Hirano and P.~Ouyang,
  ``D-brane Charges in Gravitational Duals of 2+1 Dimensional Gauge Theories
  and Duality Cascades,''
  JHEP {\bf 1001}, 072 (2010)
  [arXiv:0906.2390 [hep-th]].

  
\bibitem{Klemm:2012ii} 
  A.~Klemm, M.~Mari\~no, M.~Schiereck and M.~Soroush,
  ``Aharony-Bergman-Jafferis-Maldacena Wilson loops in the Fermi gas approach,''
  Z.\ Naturforsch.\ A {\bf 68}, 178 (2013)
  doi:10.5560/ZNA.2012-0118
  [arXiv:1207.0611 [hep-th]].
  
\bibitem{Maldacena:1998im} 
  J.~M.~Maldacena,
  ``Wilson loops in large $N$ field theories,''
  Phys.\ Rev.\ Lett.\  {\bf 80}, 4859 (1998)
  doi:10.1103/PhysRevLett.80.4859
  [hep-th/9803002].
  
\bibitem{Rey:1998ik} 
  S.~J.~Rey and J.~T.~Yee,
  ``Macroscopic strings as heavy quarks in large $N$ gauge theory and anti-de Sitter supergravity,''
  Eur.\ Phys.\ J.\ C {\bf 22}, 379 (2001)
  doi:10.1007/s100520100799
  [hep-th/9803001].
    
\bibitem{Drukker:2008zx} 
  N.~Drukker, J.~Plefka and D.~Young,
  ``Wilson loops in 3-dimensional ${\cal N}=6$ supersymmetric Chern-Simons Theory and their string theory duals,''
  JHEP {\bf 0811}, 019 (2008)
  doi:10.1088/1126-6708/2008/11/019
  [arXiv:0809.2787 [hep-th]].
  
\bibitem{deBoer:1999tgo} 
  J.~de Boer, E.~P.~Verlinde and H.~L.~Verlinde,
  ``On the holographic renormalization group,''
  JHEP {\bf 0008}, 003 (2000)
  doi:10.1088/1126-6708/2000/08/003
  [hep-th/9912012].
  
\bibitem{McGough:2016lol} 
  L.~McGough, M.~Mezei and H.~Verlinde,
  ``Moving the CFT into the bulk with $T\bar T$,''
  arXiv:1611.03470 [hep-th].
  
\bibitem{Jafferis:2011zi} 
  D.~L.~Jafferis, I.~R.~Klebanov, S.~S.~Pufu and B.~R.~Safdi,
  ``Towards the F-Theorem: N=2 Field Theories on the Three-Sphere,''
  JHEP {\bf 1106}, 102 (2011)
  doi:10.1007/JHEP06(2011)102
  [arXiv:1103.1181 [hep-th]].
  
\bibitem{Polchinski:2001tt} 
  J.~Polchinski and M.~J.~Strassler,
  ``Hard scattering and gauge / string duality,''
  Phys.\ Rev.\ Lett.\  {\bf 88}, 031601 (2002)
  doi:10.1103/PhysRevLett.88.031601
  [hep-th/0109174].
  
\bibitem{Erlich:2005qh} 
  J.~Erlich, E.~Katz, D.~T.~Son and M.~A.~Stephanov,
  ``QCD and a holographic model of hadrons,''
  Phys.\ Rev.\ Lett.\  {\bf 95}, 261602 (2005)
  doi:10.1103/PhysRevLett.95.261602
  [hep-ph/0501128].
  
\bibitem{Ryu:2006bv} 
  S.~Ryu and T.~Takayanagi,
  ``Holographic derivation of entanglement entropy from AdS/CFT,''
  Phys.\ Rev.\ Lett.\  {\bf 96}, 181602 (2006)
  doi:10.1103/PhysRevLett.96.181602
  [hep-th/0603001].
  
\bibitem{Faulkner:2013ana} 
  T.~Faulkner, A.~Lewkowycz and J.~Maldacena,
  ``Quantum corrections to holographic entanglement entropy,''
  JHEP {\bf 1311}, 074 (2013)
  doi:10.1007/JHEP11(2013)074
  [arXiv:1307.2892 [hep-th]].
  
\bibitem{Benini:2015eyy} 
  F.~Benini, K.~Hristov and A.~Zaffaroni,
  ``Black hole microstates in AdS$_{4}$ from supersymmetric localization,''
  JHEP {\bf 1605}, 054 (2016)
  doi:10.1007/JHEP05(2016)054
  [arXiv:1511.04085 [hep-th]].

\bibitem{Hijano:2015zsa} 
  E.~Hijano, P.~Kraus, E.~Perlmutter and R.~Snively,
  ``Witten Diagrams Revisited: The AdS Geometry of Conformal Blocks,''
  JHEP {\bf 1601}, 146 (2016)
  doi:10.1007/JHEP01(2016)146
  [arXiv:1508.00501 [hep-th]].
 
\bibitem{Bershadsky:1993ta} 
  M.~Bershadsky, S.~Cecotti, H.~Ooguri and C.~Vafa,
  ``Holomorphic anomalies in topological field theories,''
  Nucl.\ Phys.\ B {\bf 405}, 279 (1993)
  [AMS/IP Stud.\ Adv.\ Math.\  {\bf 1}, 655 (1996)]
  doi:10.1016/0550-3213(93)90548-4
  [hep-th/9302103].
 
\bibitem{Witten:1993ed} 
  E.~Witten,
  ``Quantum background independence in string theory,''
  Salamfest 1993:0257-275
  [hep-th/9306122].
  
\bibitem{Maldacena:2002vr} 
  J.~M.~Maldacena,
  ``Non-Gaussian features of primordial fluctuations in single field inflationary models,''
  JHEP {\bf 0305}, 013 (2003)
  doi:10.1088/1126-6708/2003/05/013
  [astro-ph/0210603].
  
\bibitem{Maldacena:2011mk} 
  J.~Maldacena,
  ``Einstein Gravity from Conformal Gravity,''
  arXiv:1105.5632 [hep-th].
  

\end{thebibliography}
\end{document}